\documentclass[12pt]{article}
\usepackage{epsfig,amsmath,amsfonts,amssymb,amscd}
\usepackage{graphicx}

\newtheorem{theo}{Theorem}

\newtheorem{prop}{Proposition}[section]

\makeatletter \@addtoreset{equation}{section} \makeatother

\newcommand{\mC}{\mathbb{C}}

\newcommand{\mR}{\mathbb{R}}

\newcommand{\calL}{{\cal L}}
\newcommand{\calM}{{\cal M}}

\newcommand{\calP}{{\cal P}}

\newcommand{\thet}{\vartheta}

\newcommand{\diag}{\operatorname{diag}}

\newcommand{\re}{\operatorname{Re}}

\newcommand{\spec}{\operatorname{spec}}

\newcommand\qed{{\unskip\nobreak\hfil\penalty50
  \hskip2em\hbox{}\nobreak\hfil\mbox{\rule{1ex}{1ex} \qquad}
    \parfillskip=0pt \finalhyphendemerits=0\par\medskip}}

\begin{document}
\title{Waves, structures, and the Riemann problem for a system of hyperbolic conservation laws  \\[-1ex]}
\author{A.~P.~Chugainova, D.~V. Treschev \\Steklov Mathematical Institute of Russian Academy of Sciences,\\
 Moscow, Russia}
\date{}
\maketitle
\paragraph{Abstract}
A system of hyperbolic conservation laws
$$
\partial_t u + \partial_x \partial_u Q = 0, \quad
  Q = u_1^3 / 3 + u_1 u_2^2, \qquad
  u = u(x,t) \in\mR^2,
$$
as well as its viscous regularization
$$
 \partial_t u + \partial_x \partial_u Q = \calM \partial_x^2 u, \qquad
  \calM = \diag (\mu_1,\mu_2), \quad \mu_1>0,\, \mu_2>0,
$$
are studied. It is assumed that admissible shocks are those that
satisfy the requirement of existence of a structure (the traveling
wave criterion). A solution of the Riemann problem is constructed
that consists of rarefaction waves and shocks with structure.
Depending on the conditions imposed at $\pm\infty$, the solution
also contains undercompressive shocks and Jouguet waves.

\section{Introduction}

Consider a system of two nonlinear partial differential equations
\begin{equation}
\label{hyp}
  \partial_t u + \partial_x \partial_u Q = 0, \quad
  Q = Q(u), \qquad
  u = u(x,t) \in\mR^2.
\end{equation}
The eigenvalues $c_1\le c_2$ of the $2\times2$ matrix
$\partial^2_u Q(u)$ are the characteristic velocities at the point
$u$. Since the matrix $\partial^2_u Q$ is symmetric, the
eigenvalues are always real; therefore, system (\ref{hyp}) is
hyperbolic. The strict hyperbolicity condition $c_1\ne c_2$ may
not be satisfied at some points.

 Discontinuous solutions of the form
\begin{equation}
\label{disc}
  u = u(\xi) = \left\{\begin{array}{cc}  u^+ & \mbox{if } \xi > 0, \\
                                         u^- & \mbox{if } \xi < 0,
                      \end{array},
               \right. \qquad
  \xi = x - Wt,
\end{equation}
where $W$ is the shock velocity, are called shocks.

The existence of solution (\ref{disc}) requires that the constants
$u^\pm$ lie on the same Hugoniot locus (Rankine--Hugoniot locus).
The Hugoniot locus is defined by the point $u^+$ and is given by
the conservation laws at the discontinuity:
\begin{equation}
\label{RH0}
  \{u\in\mR^2 : W(u-u^+) = \partial_u Q(u) - \partial_u Q(u^+)\} .
\end{equation}
However, not all the points $u=u^-$ of the Hugoniot locus define a
shock. We should also require that the evolutionary condition (the
Lax condition \cite{Lax}) be satisfied. For the $2\times2$ system,
these conditions are given by the following inequalities:
\begin{eqnarray}
\label{ev_b}
&  c_2^+ < W, \quad c_1^- < W < c_2^- \quad
   \mbox{fast compressive shocks},  & \\
\label{ev_m}
&  c_1^+ < W < c_2^+, \quad W < c_1^- \quad
   \mbox{slow compressive shocks}. &
\end{eqnarray}
Here $c_{1,2}^\pm$ are the characteristic velocities ahead of
$(+)$ and behind $(-)$ the shock.

A special role is played by undercompressive and overcompressive
shocks \cite{Dafermos} (nonclassical shocks). The velocity of
undercompressive shocks satisfies inequalities (\ref{under}),
while the velocity of overcompressive shocks satisfies
inequalities (\ref{over}):
\begin{eqnarray}
\label{under}
&   c_1^+ < W < c_2^+, \quad c_1^- < W < c_2^- , & \\
\label{over}
&   c_2^+ < W < c_1^- . &
\end{eqnarray}

The functions (\ref{disc}) are solutions of equations (\ref{hyp})
only in the generalized sense. It is well known (see, for example,
\cite{Dafermos}) that classical solutions of nonlinear hyperbolic
equations exist in general only on a finite time interval (the
gradient catastrophe). Therefore, along with system (\ref{hyp}),
we also consider it regularization
\begin{equation}
\label{reg}
  \partial_t u + \partial_x \partial_u Q = \calM \partial_x^2 u, \qquad
  \calM = \diag (\mu_1,\mu_2).
\end{equation}
Here $\mu_1$ and $\mu_2$ are positive constants. The right-hand
side of (\ref{reg}) has the physical meaning of dissipative terms.
In applications these terms are often considered to be small,
although the key role in qualitative analysis is played not by the
values of the parameters $\mu_1$ and $\mu_2$, but by their ratio,
since the system (\ref{reg}) is invariant under rescaling
$$
  t\mapsto st, \quad
  x\mapsto sx, \quad
  \mu_1\mapsto s \mu_1, \quad
  \mu_2\mapsto s \mu_2.
$$

For system (\ref{reg}) we can consider traveling wave solutions
\begin{equation}
\label{tws}
  u = u(\xi), \qquad  \xi = x - Wt, \quad
  \lim_{\xi\to\pm\infty} u(\xi) = u^\pm.
\end{equation}
As $\mu_1,\mu_2\to 0$, these solutions turn into discontinuous
wave solutions of equation (\ref{hyp}), and the quantities $u^-$
and $u^+$ are the limiting values of $u$ at $x=-\infty$ and
$x=\infty$, respectively. In this case, the discontinuous solution
(\ref{disc}) is said to have a structure defined by the family of
solutions (\ref{tws}) $u(\xi)=u(\xi;\mu_1,\mu_2)$.

It is natural to take the above-mentioned limit as $\mu_1,\mu_2
\searrow 0$ along some preselected ray on which $\mu_1/\mu_2 =
$const. Then the presence or absence of structure of a shock
depends on the ratio $\mu_1/\mu_2$. Undercompressive and
overcompressive shocks may also have a structure. In the case of
undercompressive shocks, this occurs for a special choice of the
parameters $u^+,W,\mu_1$ and $\mu_2$; more precisely, for the
values of parameters from some subset of codimension 1.

The dynamic stability of a shock is an important aspect
determining its observability. By stability we mean spectral
stability, or in other words, stability in linear approximation.

As a rule, the calculation of the spectral characteristics of the
wave solution (\ref{tws}) is a difficult problem. One of popular
methods for analyzing spectral stability is the application of the
argument principle to the Evans function \cite{Pego, Alexander};
however, this usually can be done only numerically.

In a number of publications \cite{Schaeffer, Schaeffer1, Shearer,
Shearer1, Schecter}, the authors studied the solutions of a
$2\times2$ nonstrictly hyperbolic system of conservation laws. In
these publications, the corresponding normal form is given by $Q =
pu_1^3 / 3 + qu_1^2 u_2 + u_1 u_2^2$. In \cite{Schaeffer1,
Shearer}, the authors studied the Riemann problem for equations
(\ref{hyp}) and (\ref{reg}) with such $Q$ in the case of
$\mu_1=\mu_2$.

Let $q=0$. The study of such a system is motivated by the problem
of long small-amplitude longitudinal-torsional waves propagating
in the positive direction of the $x$ axis in nonlinearly elastic
rods \cite{umn_R,Chu_2024}. The variables $u_1$ and $u_2$
characterize the deformations of the rod, and $x$ is the
Lagrangian coordinate along the axis of the rod. The deformations
$u$ are assumed small. The constant $p$ describes the elastic
properties of the medium. Nonlinearity is taken into account only
in the lowest order terms of the function $Q$ in amplitude.

Next, we consider the case of $p=1$:
\begin{equation}
\label{Q=}
  Q = u_1^3 / 3 + u_1 u_2^2.
\end{equation}
Note that system (\ref{hyp}) with potential (\ref{Q=}) is strongly
degenerate. We mean the fact that the change of variables
\begin{equation}
\label{change}
  \tilde u_1 = u_1 + u_2, \quad
  \tilde u_2 = u_1 - u_2
\end{equation}
results in two independent Hopf equations
$$
  \partial_t\tilde u_1 + 2\tilde u_1 \partial_x\tilde u_1 = 0, \quad
  \partial_t\tilde u_2 + 2\tilde u_2 \partial_x\tilde u_2 = 0.
$$
However, the problem becomes nontrivial in the context of viscous
regularization (\ref{reg}) provided that the operator $\calM$
($\mu_1\ne\mu_2$) is anisotropic.

In this paper we carry out the most comprehensive study of
discontinuous solutions for potential (\ref{Q=}) in equations
(\ref{hyp}) and the structures of these solutions defined by
equation (\ref{reg}), including

\begin{itemize}
\item   the explicit form of the Hugoniot locus for arbitrary
values of $u^+\in\mR^2$ (Section \ref{sec:RH});

\item   selecting regions on the Hugoniot locus that correspond to
fast and slow shocks (Section \ref{sec:Lax}), rarefaction waves
(Section \ref{sec:rarefaction}), undercompressive shocks (Section
\ref{sec:ss}), overcompressive shocks (Section \ref{sec:over}), as
well as Jouguet waves (Section \ref{sec:Jouguet}) and special
rarefaction waves (Section \ref{sec:special});

\item   the calculation of the parameter values corresponding to
undercompressive shocks (Section \ref{sec:ss});

\item   the solution of the problem of the presence of structure
for a shock, depending on the values of the parameters
$u^+,W,\mu_1,\mu_2$ (Sections
\ref{sec:slow_str}--\ref{sec:fast_str});

\item   the proof of the stability of the shock structures
(\ref{tws}) in the case of $\mu_1=\mu_2 > 0$ (Section
\ref{sec:stab});

\item   the solution of the Riemann problem of the decay of a
discontinuity (Section \ref{sec:Rie}).
\end{itemize}

We hope that a similar program can largely be implemented in the
case of an arbitrary $p$; but this will be the subject of another
study.

\section{The Hugoniot locus}
\label{sec:RH}

Solutions of the form (\ref{tws}) satisfy the equation
$$
  (- Wu + \partial_u Q)' = \calM u'', \qquad
  (\cdot)' = d/d\xi,
$$
or after integration with respect to $\xi$,
\begin{equation}
\label{ode}
   u'  =  - \calM^{-1} \partial_u Z(u), \qquad
 Z(u)  =  - Q + W(u_1^2 + u_2^2) / 2 + D_1 u_1 + D_2 u_2.
\end{equation}
The ordinary differential equation (\ref{ode}) is a gradient-like
system (in renormalized coordinates, gradient system) in the
plane. The coefficients $D_1$ and $D_2$ are constants. The
function $Z$ decreases along integral curves. In the Morse--Smale
theory, functions of the type $Z$ are called energy functions
(see, for example, \cite{Pochinka}).

The singular points (equilibrium states) of system (\ref{ode})
satisfy the equations $\partial_u Z = 0$:
\begin{equation}
\label{dZ=0}
  -u_1^2 - u_2^2 + Wu_1 + D_1 = 0, \quad
  -2 u_1 u_2 + Wu_2 + D_2 = 0.
\end{equation}
Hence we obtain
\begin{equation}
\label{D}
  D_1 = {u_1^+}^2 + {u_2^+}^2 - W u_1^+, \quad
  D_2 = 2 u_1^+ u_2^+ - W u_2^+ .
\end{equation}

If the determinant of the Hesse matrix
\begin{equation}
\label{hess}
  \det\partial^2_u Z = (-2u_1 + W)^2 - 4u_2^2,
\end{equation}
is positive at a singular point $u$, then $u$ is a node, while, if
it is negative, then $u$ is a saddle.

Henceforth we assume that, for any $W$, the quantities
$D_1=D_1(W,u^+)$ and $D_2=D_2(W,u^+)$ satisfy equalities
(\ref{D}), so that for all $W$ the point $u^+$ is an equilibrium
state of system (\ref{ode}).

For any $W\in\mR$, this system (\ref{dZ=0}), (\ref{D}) has four
solutions with respect to $u$ (including multiplicities):
\begin{equation}
\label{4cp}
\begin{array}{cllrcl}
  u^+ &=& (u_1^+,u_2^+), & \quad
  u^\times &=& (W-u_1^+,-u_2^+), \\
  u^a &=& (u_2^+ + W/2,u_1^+ - W/2), & \quad
  u^b &=& (W/2 - u_2^+,W/2 - u_1^+).
\end{array}
\end{equation}

\begin{prop}
\label{prop:4points} If all points (\ref{4cp}) are different, then
$u^+$ and $u^\times$ are of the same type (an extremum or a
saddle). The points $u^a$ and $u^b$ are also of the same
(opposite) type (a saddle or an extremum).
\end{prop}

{\it Proof}. It suffices to calculate (\ref{hess}) at the points
(\ref{4cp}). \qed

The curve
$$
  \{u\in\mR^2 : \partial_u Z(u) = \partial_u Z(u^+), \; W\in\mR\}
$$
is called the Hugoniot locus (the Rankine--Hugoniot locus). It
consists of the equilibrium states of system
(\ref{ode})$|_{D=D(W,u^+)}$.

There is the following equivalent definition of the Hugoniot locus
in the theory of hyperbolic equations. The system of equations
(\ref{hyp}) expresses conservation laws, which correspond to the
relations at the discontinuities of the solutions $u(\xi)$:
\begin{equation}
\label{RH}
  W (u^+ - u^-) = \partial_u Q(u^+) - \partial_u Q(u^-).
\end{equation}
Here $W$ is the velocity of the discontinuity. If we fix a value
of $u^+$ and take $u = u^-$, then (\ref{RH}) gives the equation
\begin{equation}
\label{RH+}
  W(u_1^+ - u_1) = {u_1^+}^2 + {u_2^+}^2 - u_1^2 - u_2^2, \quad
  W(u_2^+ - u_2) = 2 u_1^+ u_2^+ - 2 u_1 u_2,
\end{equation}
which is equivalent to (\ref{dZ=0}).

\begin{figure}[h]
\centerline{\psfig{figure=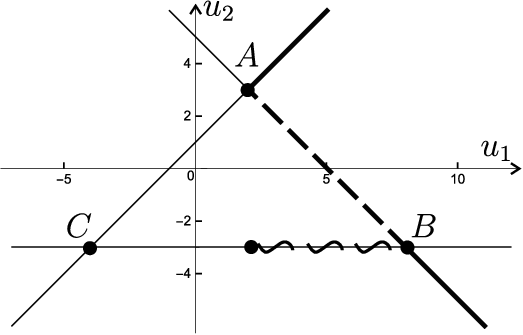,height=5cm}} \caption{Hugoniot
locus for the parameters $u_1^+ = 2$, $u_2^+ = 3$, and $p=1$.}
\label{HL}
\end{figure}

Eliminating the velocity $W$ in equalities (\ref{RH+}), we obtain
another equation of the Hugoniot locus:
\begin{equation}\label{qu333}
  (u_2 + u_2^+)(u_2 - u_1+u_1^+ - u^+_2)(u_2 + u_1 - u_1^+ - u_2^+) = 0.
\end{equation}
Therefore the Hugoniot locus represents three straight lines (see
Fig. \ref{HL})
\begin{eqnarray}
\label{first}
   u_2 &=& -u_2^+, \qquad\qquad\quad\;     W = u_1+u_1^+, \\
\label{second}
   u_2 &=& u_1 - u_1^+ + u_2^+, \quad\;\;  W = 2(u_1 + u_2^+), \\
\label{third}
   u_2 &=& -u_1 + u_1^+ + u_2^+, \quad     W = 2(u_1 - u_2^+) .
\end{eqnarray}

Let us briefly describe some details of Fig. \ref{HL}. The details
and the description of terms will be given in Sections
\ref{sec:Lax}--\ref{sec:special}. We begin with the case of $u_2^+
> 0$.

The point $A$ denotes the state $u^+$. The points of intersection
of the line (\ref{first}) with the lines (\ref{second}) and
(\ref{third}) are denoted by $C$ and $B$, respectively. On the
interval $AB$, there are points $u^-$ that define slow shocks, and
on the ray going upward right from the point $A$ and downward
right from $B$, there are points $u^-$ that define fast shocks.
The point $u^- = B$ defines a (degenerate) Jouguet shock. The ray
directed downward left from $A$ corresponds to fast rarefaction
waves, and the ray directed upward left from $A$, to slow
rarefaction waves. For certain values of the parameter
$\mu_1/\mu_2$, the points $u^-$ of the interval chosen on $CB$ may
correspond to undercompressive shocks, and the points of the ray
with origin at $B$ and directed to the right along the line
$u_2=-u_2^+$, to overcompressive shocks.

If $u_2^+ < 0$, then the Hugoniot locus overturns symmetrically
with respect to the axis $u_1$. The descriptions of the types of
waves for different points $u^-$ of the Hugoniot locus also change
symmetrically.

The case of $u_2^+ = 0$ deserves special comment. In this
situation all three lines (\ref{first})--(\ref{third}) pass
through $u^+$, so that the points $A$, $B$, and $C$ coincide. Here
there is no point in speaking of fast and slow waves. The rays
directed upward right and downward right from the point $A=u^+$
correspond to shocks, while the rays directed upward left and
downward left correspond to rarefaction waves. To the left of the
point $A$ there goes a ray consisting of the points $u^-$ defining
special rarefaction waves, whose existence is due to the
degeneracy of the problem.

\section{Evolutionary conditions for discontinuities. Shocks}
\label{sec:Lax}

Among the set of discontinuities in the solutions of the system of
equations (\ref{hyp}), we distinguish evolutionary discontinuities
(shocks), which satisfy the Lax conditions \cite{Lax}. The Lax
conditions are expressed as inequalities
(\ref{ev_b})--(\ref{ev_m}) between the shock velocity $W$ and the
characteristic velocities $c_1^+ < c_2^+$ ahead of and $c_1^- <
c_2^-$ behind the shock. The evolutionary conditions
(\ref{ev_b})--(\ref{ev_m}) are equivalently expressed as
\begin{eqnarray}
\label{ev_b1}
&  \lambda_2^+ < 0, \quad \lambda_1^- < 0 < \lambda_2^- \quad
   \mbox{fast shocks},  & \\
\label{ev_m1}
&  \lambda_1^+ < 0 < \lambda_2^+, \quad 0 < \lambda_1^-  \quad
   \mbox{slow shocks} , &
\end{eqnarray}
where $\lambda_j^\pm = c_j^\pm - W$ are the eigenvalues of the
matrix $-\partial^2_u Z(u^\pm)$. Note that, generally
speaking,\footnote {if the operator $\calM$ is not scalar} the
characteristic numbers of singular points of equation (\ref{ode}))
do not coincide with $\lambda_1$ and $\lambda_2$ but are closely
related to it: a singular point is a node if $\lambda_1 \lambda_2
> 0$ and a saddle if $\lambda_1\lambda_2<0$.

Conditions (\ref{ev_b1})--(\ref{ev_m1}) imply that the type of the
critical points $u^+$ and $u=u^-$ of the function $Z$ is as
follows: either $u$ is a saddle and $u^+$ is a local minimum, or
$u$ is a local maximum and $u^+$ is a saddle. Since the type of a
singular point is determined by the sign of expression
(\ref{hess}), we obtain the evolutionary condition in the form

\begin{equation}
\label{heshes}
   \Big((-u_1 + W/2)^2 - u_2^2\Big) \Big((-u_1^+ + W/2)^2 - {u_2^+}^2\Big) < 0;  \\
\end{equation}
in this case, (\ref{ev_b1})--(\ref{ev_m1}) give an additional
condition:
\begin{eqnarray*}
\label{slow}
&  \mbox{either the matrix $\partial_u^2 Z(u^+)$ is positive definite (fast wave)}, & \\
\label{fast} &  \mbox{or the matrix $\partial_u^2 Z(u)$ is
negative definite (slow wave)}. &
\end{eqnarray*}
Let us verify if the evolutionary conditions hold at the points of
the Hugoniot locus. Recall that the quantity $u_2^+$ is assumed to
be nonnegative.

\begin{prop}
\label{prop:slowfast} The states $u^-$ corresponding to slow
shocks lie in the interval $AB$, and the values of $u^-$
corresponding to fast shocks lie on the straight line $CA$ to the
right of the point $A$ and on the line $AB$ to the right of the
point $B$.
\end{prop}

{\it Proof}. On the component (\ref{first}) of the Hugoniot locus,
inequality (\ref{heshes}) takes the form
$$
  \Big( (u_1 - u_1^+)^2 / 4 - {u_2^+}^2 \Big)^2 < 0.
$$
It does not hold at any point of the line (\ref{first}).

On the component (\ref{second}), inequality (\ref{heshes}) turns
out to be as follows:
$$
  - \Big( {u_2^+}^2 - (u_1 - u_1^+ + u_2^+)^2 \Big)^2 < 0.
$$
It holds always (if $u_1 - u_1^+ + u_2^+ \ne \pm u_2^+$). Since
\begin{eqnarray*}
     \partial^2_u Z(u)
 &=& 2\left(\begin{array}{cc} u_2^+             & -u_1+u_1^+-u_2^+ \\
                            -u_1+u_1^+-u_2^+   & u_2^+
           \end{array}
      \right), \\
      \partial^2_u Z(u^+)
 &=& 2\left(\begin{array}{cc} -u_1^+ + u_1 + u_2^+ & -u_2^+ \\
                               -u_2^+              & -u_1^+ + u_1 + u_2^+
            \end{array}
      \right),
\end{eqnarray*}
we find that fast waves correspond to the interval $0 < u_1 -
u_1^+$ and that there are no slow waves.

Consider the straight line (\ref{third}). Inequality
(\ref{heshes}) takes the form
$$
  - \Big( ({u_2^+}^2 - (u_1 - u_1^+ - u_2^+)^2 \Big)^2 < 0.
$$
It holds always (if $u_1 - u_1^+ - u_2^+\ne \pm u_2^+$). Since
\begin{eqnarray*}
     \partial^2_u Z(u)
 &=& 2\left(\begin{array}{cc} -u_2^+             & u_1 - u_1^+ - u_2^+ \\
                            u_1 - u_1^+ - u_2^+  & - u_2^+
           \end{array}
      \right), \\
      \partial^2_u Z(u^+)
 &=& 2\left(\begin{array}{cc} u_1 - u_1^+ - u_2^+ & -u_2^+ \\
                               -u_2^+             & u_1 - u_1^+ - u_2^+
            \end{array}
      \right),
\end{eqnarray*}
we find that fast waves on the line (\ref{third}) correspond to
points at which $2u_2^+ < u_1 - u_1^+$. Slow waves correspond to
the interval $0 < u_1-u_1^+ < 2u_2^+$. \qed

The result of these calculations is illustrated in Fig. \ref{HL}.

\section{Rarefaction waves}
\label{sec:rarefaction}

Rarefaction waves in equation (\ref{hyp}) are solutions of the
form $u = u(\thet)$, $\thet = x/t$. Substitution into (\ref{hyp})
yields a system of ordinary differential equations
\begin{equation}\label{q93}
    (\partial_u^2 Q - \thet) \partial_\thet u = 0.
\end{equation}
The nontrivial solution $\partial_\thet u$ of the linear
homogeneous system of equations (\ref{q93}) exists if $\thet$ is
an eigenvalue of the linear operator $\partial_u^2 Q(u(\thet))$,
in other words, if it is the characteristic velocity.

Assuming that $u_2 > 0$, at every point $u\in\mR^2$ we obtain the
eigenvalues $\thet_s < \thet_f$ and eigenvectors $\beta_s,\beta_f$
corresponding to slow and fast rarefaction waves:
$$
  \thet_s = 2 u_1 - 2u_2, \quad
  \beta_s = \bigg(\!\!\begin{array}{c} 1\\ -1 \end{array}\!\!\bigg), \qquad
  \thet_f = 2 u_1 + 2 u_2, \quad
  \beta_f = \bigg(\!\begin{array}{c} 1\\ 1 \end{array}\!\bigg).
$$

Self-similar solutions corresponding to rarefaction waves have the
form
\begin{eqnarray*}
&&\!\!\!
   \thet_s = \thet, \quad
             2u_1 + 2u_2 = \mbox{const}. \quad
   \mbox{(slow rarefaction waves)}, \\
&&\!\!\!
   \thet_f = \thet, \quad
             2u_1 - 2u_2 =  \mbox{const}. \quad
   \mbox{(fast rarefaction waves)}.
   \end{eqnarray*}

Thus, rarefaction waves exist when
\begin{eqnarray}
\label{fast_rare}
  u^+ - u^- = \lambda\Big(\!\begin{array}{c} 1\\ 1 \end{array}\!\Big), \quad \mbox{(fast waves)}, \\
\label{slow_rare}
  u^+ - u^- = \lambda \Big(\!\begin{array}{c} 1\\ -1 \end{array}\!\Big)  \quad \mbox{(slow waves)}
\end{eqnarray}
($\lambda > 0$ is an arbitrary constant) and have the form
\begin{eqnarray*}
      u
   =  u_s(\thet)
  &=& \left\{ \begin{array}{cl} u^-  &\mbox{if }  \thet < 2u_1^- - 2u_2^- ,  \\
                                \displaystyle
                                \frac{\thet}{4}\Big(\!\begin{array}{c} 1\\ -1 \end{array}\!\Big)
                                + \frac{u_1^+ + u_2^+}{2}\Big(\!\begin{array}{c} 1\\ 1 \end{array}\!\Big)
                                     &\mbox{if }  2u_1^- - 2u_2^-  < \thet < 2u_1^+ - 2u_2^+ , \\
                                u^+  &\mbox{if }  2u_1^+ - 2u_2^+ < \thet ,
              \end{array}
      \right. \\
      u
   =  u_f(\thet)
  &=& \left\{ \begin{array}{cl} u^-  &\mbox{if }  \thet < 2u_1^- + 2u_2^- ,  \\
                                \displaystyle
                                \frac{\thet}{4}\Big(\!\begin{array}{c} 1\\ 1 \end{array}\!\Big)
                                + \frac{u_1^+ - u_2^+}{2}\Big(\!\begin{array}{c} 1\\ -1 \end{array}\!\Big)
                                     &\mbox{if }  2u_1^- + 2u_2^-  < \thet < 2u_1^+ + 2u_2^+ , \\
                                u^+  &\mbox{if }  2u_1^+ + 2u_2^+ < \thet .
              \end{array}
      \right.
\end{eqnarray*}

Particles in the rarefaction wave move with different velocities.
This velocity monotonically changes from the left end (point
$u^-$) to the right end (point $u^+$), which move with velocities
\begin{eqnarray}
\label{Ws}
   c_s(u^-) = 2u_1^- - 2u_2^-, \quad  c_s(u^+) = 2u_1^+ - 2u_2^+, \\
\label{Wf}
   c_f(u^-) = 2u_1^- + 2u_2^-, \quad  c_f(u^+) = 2u_1^+ + 2u_2^+ .
\end{eqnarray}
In the case of $u_2^+ < 0$, the subscripts $s$ and $f$ in these
equalities should be interchanged.

Rarefaction waves are defined by continuous (moreover, piecewise
smooth) functions $u_s(\thet)$ and $ u_f(\thet)$ that can be
obtained as the limits as $\mu_1,\mu_2\rightarrow 0$ of the smooth
solutions $u(x,t,\mu_1,\mu_2)$,
$u(\pm\infty,t,\mu_1,\mu_2)=u^{\pm}$ of system (\ref{reg}).

\section{Undercompressive shocks and saddle connections}
\label{sec:ss}

For the system of equations (\ref{hyp}), there may exist
undercompressive (nonclassical) shocks with structure. The
structure of an undercompressive shock is represented by an
integral curve connecting two saddle points of system (\ref{ode}).
The existence condition of a saddle connection determines the
velocity $W$ of the undercompressive shock. This value of the
velocity of the undercompressive shock provides an additional
condition on the shock.

Undercompressive shocks exist when inequalities (\ref{under}) are
satisfied. Equivalently,
\begin{equation}\label{evs_s}
  \lambda_1^- < 0 < \lambda_2^-, \quad
  \lambda_1^+ < 0 < \lambda_2^+.
\end{equation}

In Fig. \ref{HL}, the points on the Hugoniot locus that satisfy
inequalities (\ref{evs_s}) belong to the selected part of the
segment $CB$.

Suppose that the undercompressive shock has a structure (a viscous
profile) in the form of a traveling wave for the system of
equations (\ref{reg}). The existence of the undercompressive shock
provides a condition for the existence of a heteroclinic curve
connecting two saddles. Set
$$
  m = \sqrt{\frac{\mu_2}{2\mu_1 - \mu_2}}, \qquad
  0 < m < 1.
$$

\begin{prop}
\label{prop:sc} Suppose that
\begin{equation}
\label{()<}
  (u_1^+ - W/2)^2 < {u_2^+}^2,
\end{equation}
so that the equilibrium states $u^+$ and $u^\times$ (see
(\ref{4cp})) are saddles. A heteroclinic saddle connection between
$u^+$ and $u^\times$ exists if and only if
\begin{equation}
\label{m<m}
    W
  = 2( u_1^+ \pm m u_2^+ ) , \qquad
  0 < \mu_2 < \mu_1.
\end{equation}
\end{prop}

{\it Proof}. It is shown in \cite{Chicone} that saddle connections
in gradient quadratic systems of ODEs on the plane are
rectilinear. Therefore, if there exists a saddle connection from
$u^+$ to $u^\times$, the vector field $v = (v_1(u),v_2(u))$,
\begin{eqnarray*}
      \mu_1 v_1(u)
  &=&  u_1^2 + u_2^2 - Wu_1 - {u_1^+}^2 - {u_2^+}^2 + W u_1^+, \\
      \mu_2 v_2(u)
  &=&  2 u_1 u_2 - Wu_2 - 2 u_1^+ u_2^+ + W u_2^+,
\end{eqnarray*}
at the points of the rectilinear segment\footnote {and even at the
points of the line passing through $u^+$ and $u^\times$}
$[u^+,u^\times]$ is parallel to this segment. We obtain the
equation
$$
    v_1\big( u_1^+ + s(W-2u_1^+), u_2^+ - 2su_2^+ \big) (-2 u_2^+)
  = v_2\big( u_1^+ + s(W-2u_1^+), u_2^+ - 2su_2^+ \big) (W-2 u_1^+),
$$
which should hold for all $s\in\mR$. Hence we obtain
$$
  W = 2 ( u_1^+ \pm m u_2^+ ).
$$
This equality ensures the existence of a rectilinear heteroclinic
connecting $u^+$ with $u^\times$. Condition (\ref{()<}) is
satisfied in the case of $0 < m < 1$, which is equivalent to the
inequality $0<\mu_2<\mu_1$. \qed

\textbf{Remark.} Under condition (\ref{m<m}) we have

\begin{equation}\label{ZZ}
  Z(u^\times)-Z(u^+)=\pm4 m {u_2^+}^3(1+m^2/3).
\end{equation}
Since the function $Z$ should decrease along the integral curve of
equation (\ref{ode}), for $u_2^+>0$ we should take only the sign
$+$ in (\ref{m<m}). In this case undercompressive shocks lie on
the selected part of the segment $BC$ of the Hugoniot locus (Fig.
\ref{HL}).

\begin{prop}
\label{prop:sc+} Suppose that
\begin{equation}
\label{()>}
  (u_1^+ - W/2)^2 > {u_2^+}^2,
\end{equation}
so that the equilibrium states $u^a$ and $u^b$ (see (\ref{4cp}))
are saddles. A heteroclinic saddle connection between $u^a$ and
$u^b$ exists if and only if
\begin{equation}
\label{<<<}
  \frac{W}2 = u_1^+ \pm m^{-1} u_2^+  , \qquad
  0 < \mu_2 < \mu_1.
\end{equation}
\end{prop}

The proof follows the same line as the proof of Proposition
\ref{prop:sc}. We omit the details. \qed

\section{Overcompressive shocks}
\label{sec:over}

On the horizontal component of the Hugoniot locus (Fig. \ref{HL})

(b) to the right of $B$ and

\begin{figure}[h]
\centerline{\psfig{figure=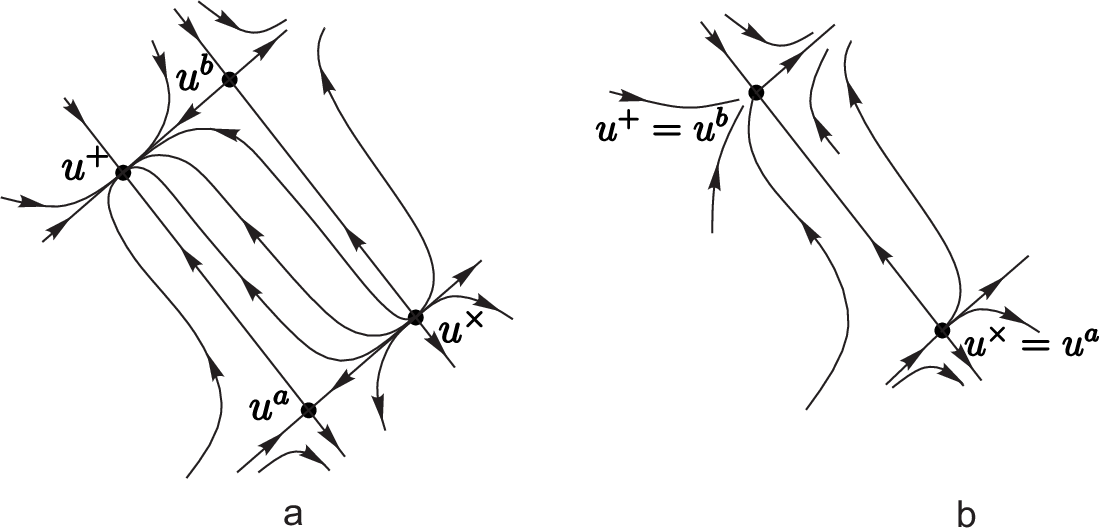,height=6cm}} \caption{(a) The
structure of (node--node) overcompressive shocks, and (b) the
saddle-node--saddle-node heteroclinic connection}.
\label{over_fig}
\end{figure}

(c) to the left of $C$,

\noindent there are points $u = u^\times$ corresponding to
saddle-type equilibrium positions in system (\ref{ode}). For
points of type (b), the function $Z$ has a maximum at $u^\times$
and a minimum at $u^+$. For points of type (c), conversely,
$u^\times$ is a minimum and $u^+$, a maximum. Therefore (since $Z$
decreases along the flow of equations (\ref{ode})), heteroclinics
from $u^\times$ to $u^+$ can exist only in the case of (b) (Fig.
\ref{over_fig}a). The heteroclinic solution (b) from $u^\times$ to
$u^+$ (if it exists) is included in the one-parameter family of
solutions of the same type. Each of these solutions may correspond
to the structure of the overcompressive shock in equation
(\ref{hyp}). Stability of overcompressive shocks is discussed for example
in \cite{Nonlinear_S_S}.

\section{Jouguet wave}
\label{sec:Jouguet}

The solutions of equations (\ref{reg}) may also include waves of
special type. These waves are called Jouguet waves. A Jouguet wave
arises when one equilibrium position, $u^+$ or $u^-$, is
degenerate in system (\ref{ode}). In this case the degenerate
point becomes a saddle-node. If the function $Q$ is defined in the
form (\ref{Q=}), then the above degeneracies occur at the points
$A$, $B$, and $C$ of the Hugoniot locus, both points $u^+$ and
$u^-$ becoming saddle-nodes. Therefore, it would be more
appropriate to call these waves degenerate Jouguet waves.

In particular, note that if a heteroclinic solution from $u^- = B
= (u_1^+ + 2u_2^+, -u_2^+)$ to $u^+ = A$ exists, then the velocity
of the corresponding Jouguet wave is $W = 2u_1^+ + 2u_2^+$ (see
(\ref{third})). The phase portrait of system (\ref{ode}) for $u^-
= B$ is demonstrated in Fig. \ref{over_fig}b.

\section{Special rarefaction waves}
\label{sec:special}

Under the conditions
\begin{equation}
\label{cond_sp}
  u_2^+ = u_2^- = 0, \quad
  u_1^- < u_1^+,
\end{equation}
there arises a special rarefaction wave. In this case equation
(\ref{reg}) takes the form
$$
  \partial_t u_1 + \partial_x (u_1^2 + u_2^2) = \mu_1\partial_x^2 u_1, \quad
  \partial_t u_2 + \partial_x (2 u_1 u_2) = \mu_2\partial_x^2 u_2.
$$
If we restrict this system to the invariant manifold $\{u_2\equiv
0\}$, then the variable $u_1$ will satisfy the Burgers equation
$\partial_t u_1 + 2u_1\partial_x u_1 = \mu_1\partial_x^2 u_1$;
under conditions (\ref{cond_sp}), we obtain solutions for this
equation that turn, as $\mu_1\to 0$, into the rarefaction wave of
the Hopf equation.

\section{Existence of a structure of a slow shock}
\label{sec:slow_str}

Henceforth we denote for short the pair $(\mu_1,\mu_2)$ by letter
$\mu$. Let

$$\calP = \{(u^+,W,\mu)\in\mR^5:\; u_2^+ \ge 0, \; \mu_1\ge 0, \; \mu_2\ge 0\}$$ be the
parameter space. Consider a region in the parameter space $\calP$
such that $u^+$ is a saddle at the points of this region:
$$
  \calP_{saddle} = \{(u^+,W,\mu)\in\calP :  2(u_1^+ - u_2^+) < W < 2(u_1^+ + u_2^+)\}.
$$
The inequality in the definition of $\calP_{saddle}$ is equivalent
to (\ref{()<}).

Let $u^- = u^a$ be a point of the Hugoniot locus that corresponds
to a  slow wave with velocity $W$. This means that the parameters
in system (\ref{ode}) lie in the region $\calP_{saddle}$ and $u^-$
is a node corresponding to a local minimum of the function $Z$.
Consider the question of existence of a structure of the shock
from $u^-$ to $u^+$.

The existence of such a structure requires the existence of a
heteroclinic solution $\gamma$ of system (\ref{ode}) going from
$u^-$ to $u^+$. The curve $\gamma$ must coincide with one of the
two branches of the stable separatrix of the saddle $u^+$.

However, the separatrix branch $\gamma$ may go to infinity, and
even go to another saddle $\hat u$, implementing a saddle
connection.\footnote {The second node corresponds to a local
minimum of the function $Z$. Therefore, $\gamma$ cannot reach this
node since $Z$ must decrease along $\gamma$.} Let
$\calP_-\subset\calP_{saddle}$ be a set of parameters for which
there exists a heteroclinic from $u^-$ to $u^+$. We also define
$\calP_\infty\subset\calP_{saddle}$ -- a set of parameters for
which both stable separatrices of the saddle $u^+$ come from
$\infty$.

A saddle connection arises in a submanifold $\calP_{\hat
u}\subset\calP_{saddle}$ of codimension 1. The submanifold
$\calP_{\hat u}$ contains the common boundary of the sets
$\calP_-$ and $\calP_\infty$:
$$
  \calP_{\hat u} \supset \partial\calP_- \cap \partial\calP_\infty .
$$
Using Proposition \ref{prop:sc}, we define open sets into which
$\calP_{\hat u}$ divides $\calP_{saddle}$:
\begin{eqnarray}
\label{set1}
\displaystyle
     u_1^+ + m u_2^+
   < &\!\! W/2 &\!\!
   < u_1^+ + u_2^+ , \qquad
     \mu_2 < \mu_1, \\
\nonumber
\displaystyle
     u_1^+ - m u_2^+
   < &\!\! W / 2 &\!\!
   < u_1^+ + m u_2^+ , \qquad
     \mu_2 < \mu_1  \\
\label{set2}
     \mbox{or} && \!\!\!\!\!\!\!\!
     u_1^+ - u_2^+ < W / 2 < u_1^+ - u_2^+, \qquad \mu_1 \le \mu_2, \\
\label{set3}
\displaystyle
     u_1^+ - u_2^+
   < &\!\! W / 2 &\!\!
   < u_1^+ - m u_2^+, \qquad
     \mu_2 < \mu_1 .
\end{eqnarray}

\begin{figure}[h]
\centerline{\psfig{figure=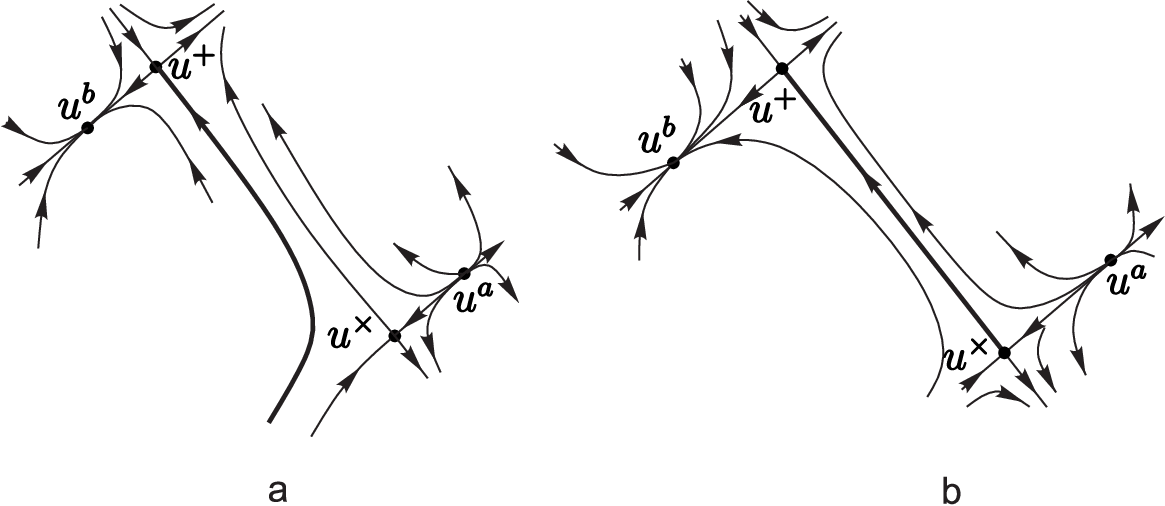,height=6cm}} \caption{The phase
portraits of system (\ref{ode}) for $\mu_2<\mu_1$: (a) parameters
lie in $\calP_\infty$, and (b) parameters lie in $\calP_{\hat u}$.
} \label{RR1}
\end{figure}

\begin{figure}[h]
\centerline{\psfig{figure=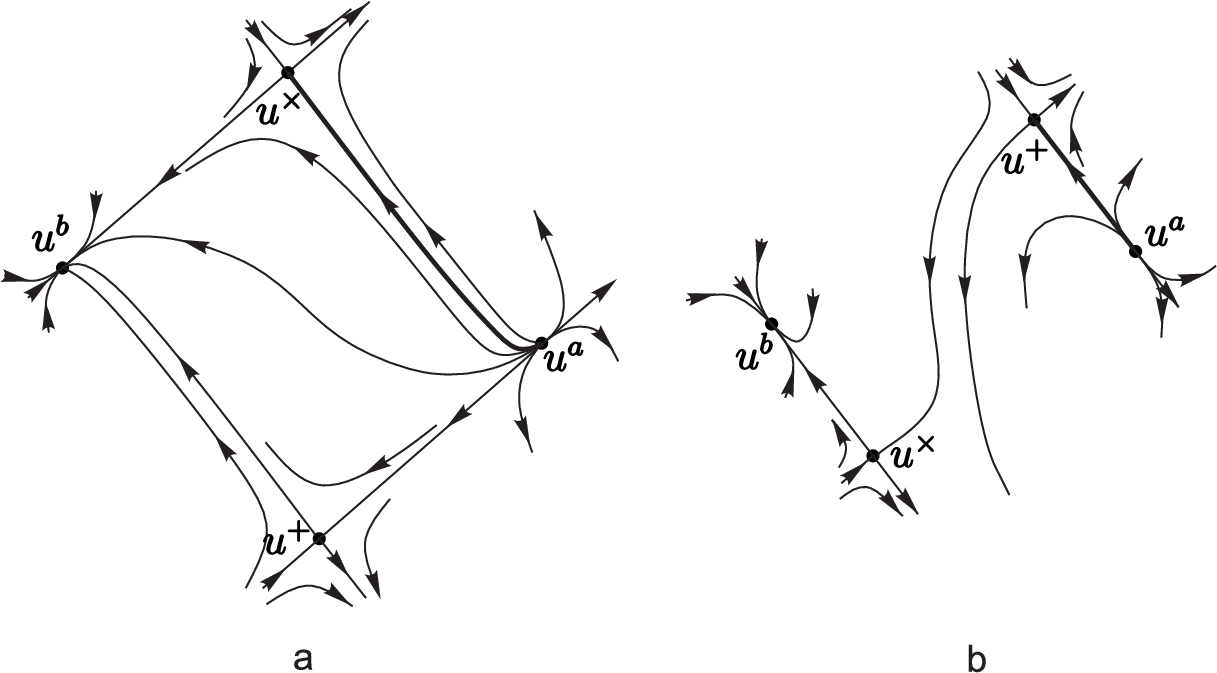,height=7cm}} \caption{The phase
portraits of system (\ref{ode}) for $\mu_2<\mu_1$ lying in
$\calP_{saddle}$: (a) in region (\ref{set2}) and (b) in region
(\ref{set3}). } \label{RR2}
\end{figure}

The phase portraits of system (\ref{ode}) for the parameter values
taken in region (\ref{set1}) and on the boundary between regions
(\ref{set1}) and (\ref{set2}) are demonstrated in Fig. \ref{RR1}.
The left part of the figure corresponds to the parameter values
satisfying inequalities (\ref{set1}). Both stable separatrices of
the saddle $u^+$ come from infinity. There is a saddle connection
in the right part of Fig. \ref{RR1}. Figure \ref{RR2} corresponds
to inequalities (\ref{set2}) and (\ref{set3}). In both cases there
exists a heteroclinic solution from $u^-=u^a$ to $u^+$. The saddle
connections separating regions (\ref{set2}) and (\ref{set3}) do
not prevent the existence of this solution.

Thus, $\calP_-$ coincides with the union of sets (\ref{set2}) and
(\ref{set3}) and includes points at which $W/2 = u_1^+ - m$. The
set $\calP_\infty$ is defined by inequalities (\ref{set1}). As a
corollary, we obtain the following two propositions.

\begin{prop}
\label{prop:D} Suppose that the inequality $0<\mu_2<\mu_1$ holds.
Then the points of the Hugoniot locus that define a slow shock
with structure lie on the segment $AB$ (see Fig. \ref{HL}) between
the points $A$ and
\begin{equation}
\label{D=()}
  D = u^+ + (1+m) u_2^+ (1, -1).
\end{equation}
The points of the segment $DB$ have no structure. For
$0<\mu_1\le\mu_2$, all slow waves have structure.
\end{prop}

\begin{prop}
\label{rem:D} In the case of $u_2^+ < 0$, equality (\ref{D=()}) is
rewritten as
$$
  D = u^+ - (1+m) u_2^+ (1, 1).
$$
\end{prop}

\section{Existence of a structure of a fast shock}
\label{sec:fast_str}

We can similarly consider the topological obstructions to the
existence of a structure of a fast shock. The question concerns
the range of parameters corresponding to the situation where $u^+$
is a local minimum of the function $Z$:
$$
  \calP_{\max} = \{(u^+,W,\mu)\in\calP : u_1^+ - W/2 < - u_2^+\}.
$$
In this case, $(W - u_1^+, -u_2^+)$ is a local maximum of $Z$, and
the equilibrium states
$$
  u^a = (u^+_2 + W/2, u_1^+ - W/2)\quad
  \mbox{and}\quad
  u^b = (W/2 - u^+_2, W/2 - u_1^+)
$$
are saddles.

The shock starts at the point
\begin{equation}
\label{fast1}
  u^- = u^a, \qquad
  W\ge 2(u_1+u_2^+)
\end{equation}
(a fast shock in the lower half-plane in Fig. \ref{HL}) or at the
point
\begin{equation}
\label{fast2}
  u^- = u^b, \qquad
  W\ge 2(u_1+u_2^+)
\end{equation}
(a fast shock in the upper half-plane in Fig. \ref{HL}).

Consider the question of existence of a structure of a fast shock.
In other words, we are interested in the existence of a
heteroclinic solution $\gamma$ from $u^-$ to $u^+$. The curve
$\gamma$ should pass along one of the branches of the unstable
separatrix of the saddle $u^-$.

Define the sets of parameters:

\smallskip
$\calP_a\subset\calP_{\max}$, for which there exists a
heteroclinic from $u^a$ to $u^+$,

$\calP_b\subset\calP_{\max}$, for which there exists a
heteroclinic from $u^b$ to $u^+$,

$\calP_\infty\subset\calP_{a\max}$, for which both branches of the
unstable separatrix of the saddle $u^a$ go to $\infty$,

$\calP_\infty\subset\calP_{b\max}$, for which both branches of the
unstable separatrix of the saddle $u^b$ go to $\infty$,

$\calP_{ab}\subset\calP_{\max}$, for which there exists a saddle
connection between $u^a$ and $u^b$.
\smallskip

The sets $\calP_a$, $\calP_b$, $\calP_{a\infty}$, and
$\calP_{b\infty}$ are open, and $\calP_{ab}$ is a submanifold of
codimension 1 defined by conditions (\ref{<<<}). The boundaries of
the sets $\partial\calP_a\cap\partial\calP_{a\infty}$ and
$\partial\calP_b\cap\partial\calP_{b\infty}$ are contained in
$\calP_{ab}$. The set $\calP_{ab}$ splits $\calP_{\max}$ into two
open connected components:
\begin{eqnarray}
\label{<W<}
    u_1^+ + u_2^+
  <\!\!  & W / 2 &\!\!
  < u_1^+ + m^{-1} u_2^+, \qquad 0 < \mu_2 < \mu_1, \\
\label{<W}
    u_1^+ + m^{-1} u_2^+
  <\!\!  & W / 2 &\!\! , \quad 0 < \mu_2 < \mu_1 \quad
    \mbox{or}\quad  0 < \mu_1 \le \mu_2.
\end{eqnarray}

\begin{figure}[h]
\centerline{\psfig{figure=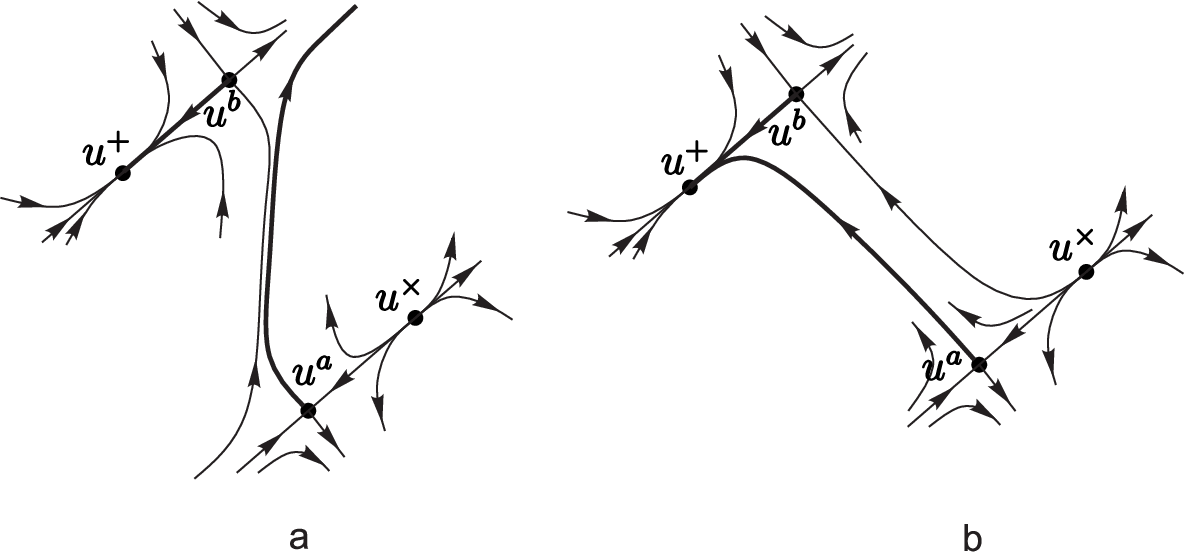,height=6cm}} \caption{The phase
portraits of system (\ref{ode}) for $\mu_2<\mu_1$ for the
parameters (a) from region (\ref{<W<})) and (b) from region
(\ref{<W})). } \label{F1}
\end{figure}

Figure \ref{F1} demonstrates the singular points and phase
portraits of system (\ref{ode}) for $\mu_2<\mu_1$. The values of
the parameter $W$ are taken on different sides of the submanifold
$\calP_{ab}$. According to Fig. \ref{F1}a, in region (\ref{<W<})
only the wave $u^b\to u^+$ has structure, while, in region
(\ref{<W}) (Fig. \ref{F1}b) both waves $u^a\to u^+$ and $u^b\to
u^+$ have structure, so that $\calP_b = \calP_{\max}$, while
$\calP_a\subset\calP_{\max}$ is defined by inequalities
(\ref{<W}). In other words, the following proposition is valid.

\begin{prop}
\label{prop:E} The fast shock $u^b\to u^+$ always has a structure.

If the inequality $0<\mu_2<\mu_1$ holds, then the points of the
Hugoniot locus that define the fast shock $u^a\to u^+$ with
structure lie on the straight line $AB$ (see Fig. \ref{HL}) below
the point
\begin{equation}
\label{E=()}
  E = u^+ + (1 + 1/m) u_2^+ (1,-1).
\end{equation}
The waves $u^a\to u^+$ corresponding to the points of the segment
$BE$ have no structure. When $0<\mu_1\le\mu_2$, all fast waves
have structure.
\end{prop}

For the parameters lying in the region $\calP_a$ (on the right
part of Fig. \ref{F1} in the phase portrait), there is a
one-parameter family of heteroclinic solutions going from the node
$\hat u$ to the node $u^+$. In the theory of discontinuous
solutions of hyperbolic equations, one speaks in this case of
overcompressive shocks.

\section{Stability of the structures of shocks}
\label{sec:stab}

Let $u^0 = u^0(\xi)$ be a heteroclinic solution of system
(\ref{ode}). It corresponds to a wave solution $u^0 = u^0(x-Wt)$
of equation (\ref{reg}). Consider the question of the dynamic
stability of this solution. In this context one usually discusses
the spectral stability of the zero solution of equation
(\ref{reg}) linearized at $u^0$. Substituting $u = u^0(\xi) +
e^{\lambda t} v(\xi)$ into (\ref{reg}) we obtain
\begin{equation}
\label{calL}
   \calL v = \lambda v, \qquad
   \calL v = \calM v'' + \big( - \partial^2_u Q(u^0) v + Wv\big)'
\end{equation}
as a first approximation in $v$. Depending on the choice of the
solution, the coefficient $W$ satisfies one of the equalities
(\ref{first})--(\ref{third}). The operator $\calL$ is considered
in the space of functions $v:\mR\to\mC$, $v\in L^2(\mR,\mC^2)$.

{\it A wave solution $u^0$ is said to be linearly stable if the
spectrum of the operator $\calL$ lies in the left
half-plane\textup: $\spec(\calL)\subset \{\lambda\in\mC:
\re\lambda\le 0\}$. If $\spec(\calL) \cap \{\lambda\in\mC:
\re\lambda > 0\} \ne\emptyset$, then $u^0$ is said to be
spectrally unstable. }

As a rule, the analysis of the spectrum of the operator $\calL$ is
a difficult problem. One usually restricts oneself to the question
of existence or absence of a discrete spectrum in the half-plane
$\{\re\lambda > 0\}$. To this end, one uses the Evans function in
combination with the argument principle \cite{Pego, Alexander}.

\subsection{The case of $\mu_1=\mu_2$}

We discuss the question of stability only in the simplest
situation of $\mu_1 = \mu_2 > 0$. In this case, the problem of
stability of the structure of a shock is solved completely.
\smallskip

In the case of $\mu_1=\mu_2$, by renormalizing the variables $x$
and $t$, we can obtain $\mu_1=\mu_2=1$, which will be used in what
follows. After the change of variables (\ref{change}), system
(\ref{reg})--(\ref{Q=}) turns into two independent Burgers
equations
\begin{equation}
\label{2Bur}
  \partial_t\tilde u_1 + 2\tilde u_1\partial_x\tilde u_1 = \partial_x^2\tilde u_1, \quad
   \partial_t\tilde u_2 + 2\tilde u_2\partial_x\tilde u_2 = \partial_x^2\tilde u_2 .
\end{equation}
The wave solution $u^-\to u^+$ takes the form
\begin{equation}
\label{2tws}
     \tilde u^-
  =  ( u_1^- + u_2^-, u_1^- - u_2^- )
 \to \tilde u^+
  =  ( u_1^+ + u_2^+, u_1^+ - u_2^+ ).
\end{equation}

\subsection{Structures of fast and slow shocks}

Let $u^-$ be a point on the Hugoniot locus that corresponds to a
slow or a fast shock, i.e.,
\begin{equation}
\label{slow-fast1}
   \mbox{either $u_1^- - u_1^+ = u_2^- - u_2^+ > 0$, or $u_1^- - u_1^+ = - u_2^- + u_2^+ > 0$}.
\end{equation}

\begin{prop}
\label{prop:stab} Suppose that $\calM = I$. Under condition
(\ref{slow-fast1}), any solution of system (\ref{reg}) with the
initial condition $u(x,0) = u_0(x)$, $\lim_{x\to\pm\infty} u_0(x)
= u^\pm$ tends asymptotically to the wave solution generated by
the heteroclinic $u^-\to u^+$ of equation (\ref{ode}).
\end{prop}

{\it Proof}. Inequalities  (\ref{slow-fast1}) turn into
\begin{equation}
\label{slow-fast2}
   \mbox{either $\tilde u_1^- > \tilde u_1^+$, $\tilde u_2^- = \tilde u_2^+$,
         or $\tilde u_1^- = \tilde u_1^+$, $\tilde u_2^- > \tilde u_2^+$}.
\end{equation}
Consider the first of these two possibilities.\footnote{The second
possibility is completely analogous.} It corresponds to the wave
$\tilde u_1^- \to \tilde u_1^+$ in the first equation (\ref{2Bur})
and the stationary solution $\tilde u_2(x,t) = \tilde u_2^- =
\tilde u_2^+$ in the second equation.

The stability of the wave solution $w^-\to w^+$, $w^+ < w^-$ to
the Burgers equation
\begin{equation}
\label{Bur}
  \partial_t w + 2 w\partial_x w = \partial_x^2 w, \qquad
  w(x,0) = w_0(x).
\end{equation}
was analyzed in \cite{IO}. In particular, the following theorem
was proved in \cite{IO}.

\begin{theo}
\label{theo:IO_1} Suppose that $w^+ < w^-$ and the initial
conditions in (\ref{Bur}) are such that there exist the following
integrals\textup:
$$
  \int_{-\infty}^0 \big( w_0(x) - w^- \big)\, dx \quad
  \mbox{and} \quad
  \int_0^{+\infty} \big( w_0(x) - w^+ \big)\, dx.
$$
Then, as $t\to +\infty$, the solution $w(x,t)$ tends uniformly
with respect to $x$ to the wave solution $\tilde w(x-Wt)$, where
$$
  W = u^+ + u^- \quad
  \mbox{and}\quad
  \int_{-\infty}^{+\infty} \big( \tilde w(s) - w_0(s) \big)\, ds = 0.
$$
Moreover, if $u_0(x)$ tends exponentially to $u^\pm$ as
$x\to\pm\infty$, then
$$
  \big| \tilde w(x-Wt) - w(x,t) \big| \le M e^{-\alpha t} \quad
  \mbox{for all $t\ge 0$ and $x$}
$$
for some constants $M,\alpha> 0$.
\end{theo}

The stability of stationary solutions of the Burgers equation
follows from another Theorem in \cite{IO}.

\begin{theo}
\label{theo:IO_2} Let $w(x,t)$ be a solution to the Cauchy problem
(\ref{Bur}). Let $\lim_{x\to\pm\infty} w_0(x) = \tilde w =$const.
Then $w(x,t)$ tends to $\tilde w$ as $t\to +\infty$ uniformly with
respect to $x$.
\end{theo}

Proposition \ref{prop:stab} follows from Theorems
\ref{theo:IO_1}--\ref{theo:IO_2}.  \qed

\subsection{Structures of overcompressive shocks}

When $\mu_1=\mu_2$, there are no saddle connections (as well as
undercompressive shocks). In Section \ref{sec:over} we have
established that the points of the Hugoniot locus corresponding to
overcompressive shocks lie on a horizontal segment to the right of
the point $B$ (Fig. \ref{HL}) and have the form $u^- = u^\times$.
However, the existence of the heteroclinic solution $u^\times\to
u^+$ requires that the additional conditions obtained in Section
\ref{sec:fast_str} should be satisfied. These conditions have the
form (\ref{<W}) together with the inequality $u_1^+ + u_2^+ <
W/2$.

With regard to (\ref{first}), in the variables $\tilde u$ we deal
with wave solutions (\ref{2tws}) of the form
\begin{equation}
\label{overcomp}
      \tilde u_1^- = W - u_1^+ - u_2^+
  \to \tilde u_1^+ = u_1^+ + u_2^+, \quad
      \tilde u_2^- = W - u_1^+ + u_2^+
  \to \tilde u_2^+ = u_1^+ - u_2^+ .
\end{equation}
If $\tilde u_1 = \gamma_1(\xi)$, $\tilde u_2 = \gamma_2(\xi)$ is
such a solution, then, for any constants $\xi_1$ and $\xi_2$,
$\tilde u_1 = \gamma_1(\xi - \xi_1)$, $\tilde u_2 = \gamma_2(\xi -
\xi_2)$ is also a solution of the form (\ref{overcomp}), so that
we have a one-parameter family of overcompressive shocks.

The velocities of the waves $\gamma_1(\xi-\xi_1)$ and
$\gamma_2(\xi-\xi_2)$ in the two Burgers equations coincide and
are equal to $W$. According to Theorem \ref{theo:IO_1}, the
corresponding solutions of equations (\ref{Bur}), and hence the
corresponding solutions of equation (\ref{reg}), are stable. Such
an extremely atypical phenomenon of total stability of
overcompressive shocks is naturally due to the degeneracy of the
problem, caused by the assumption $\mu_1=\mu_2$.

\section{Riemann problem}
\label{sec:Rie}

The Riemann problem on the decay of a discontinuity consists in
that, at the initial time $t=0$,
$$
    u(x,0)
  = \left\{\begin{array}{cc}
             u^-\;  & \mbox{for} \quad x<0 , \\
             u^+\;  & \mbox{for} \quad x>0 .
        \end{array}
    \right.
$$
It is required to find a solution to the system (\ref{hyp}) for
$t>0$.

The statement of the problem needs comments. Since we consider the
solutions of equation (\ref{hyp}) as the limits as $\calM\to 0$ of
the solutions of the regularized equation (\ref{reg}), we
construct a solution of the Riemann problem only from the waves
with structure. Recall that the limit as $\calM\to 0$ should be
considered for a fixed ratio $\mu_1/\mu_2$. The presence or
absence of structure of a discontinuous solution essentially
depends on this fact. Therefore, the parameter $\mu_1 / \mu_2$ is
explicitly involved in the statement of the Riemann problem and in
its solution. From the viewpoint of the physical meaning of the
solution, it is also important to raise the question of stability
of the structures obtained. We do not discuss this question.
However, we note that in any numerical experiment we did not
observe dynamically unstable wave solutions in the system that
arise from the heteroclinics of equation (\ref{ode}).

The solutions of the Riemann problem consist of a sequence of
rarefaction waves and shocks with structure that follow in the
order of decreasing velocity. The Hugoniot locus starting from the
initial point $A = u^+$ consists of three straight lines: $AB$,
$AC$, and $BC$ (Fig. \ref{HL}).

Let $u_2^+>0$. Introduce the following notations:

\begin{itemize}
\item $S_2$ is a fast shock from the point $A$ to a point of the
Hugoniot locus on the ray going upward right from the point $A$;

\item $\hat S_2$ is a fast shock from the point $A$ to a point on
the ray going downward right from the point $B$;

\item $S_1$ is a slow shock from the point $A$ to a point in the
interval $AB$;

\item $S$ is a (degenerate) Jouguet shock from $A$ to $B$;

\item $R_2$ is a fast rarefaction wave from the point $A$ to a
point on the ray going downward left from the point $A$;

\item $R_1$ is a slow rarefaction wave from the point $A$ to a
point on the ray going upward right from the point $A$;

\item $R$ is a special rarefaction wave;

\item $Z$ is an undercompressive shock.
\end{itemize}

If $u_2^+<0$, then the Hugoniot locus is mirror reflected with
respect to the axis $u_2=0$, and the directions of the
corresponding rays described above are changed to the opposite.

Consider the solution separately for the cases of $\mu_1\leqslant
\mu_2$ and $\mu_1>\mu_2$.

\subsection{The case of $\mu_1 \leqslant \mu_2$}
In this case all shocks have structure.

\begin{figure}[h]
\centerline{\psfig{figure=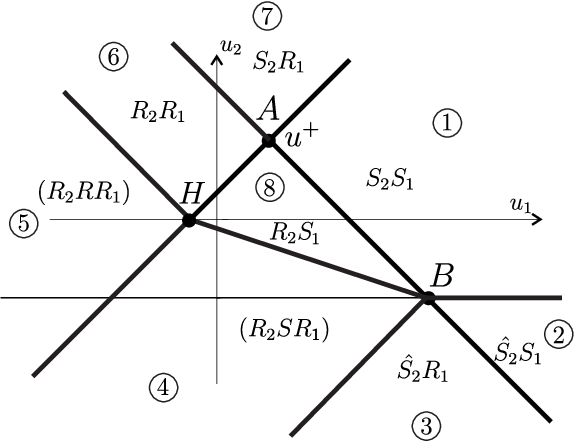,height=7cm}} \caption{Solution
of the Riemann problem in the case of $\mu_1\leqslant \mu_2$. }
\label{R11N}
\end{figure}

The solution is demonstrated in Fig. \ref{R11N}. The point $A$
represents the state to the right of the initial shock front and
has coordinates $u^+$. The state to the left of the initial shock
front is represented by an arbitrary point $u\in\mR^2$. The plane
$u$ is partitioned into regions with different configurations of
the solution.

{\bf Region 1}. We begin with the situation where the solution of
the Riemann problem consists of the fast shock $S_2$ (transition
$u\to u^+$) followed by the slow shock $S_1$ (transition $u^-\to
u$). The coordinates of the points $u$ and $u^-$ are
\begin{eqnarray}
\nonumber
      u
  &=& (u_1^+ + \lambda_+, u_2^+ + \lambda_+), \qquad \lambda_+ > 0, \\
\label{u^-(1)}
      u^-
  &=& (u_1^+ + \lambda_+ + \lambda_-, u_2^+ + \lambda_+ - \lambda_-), \qquad
      0 < \lambda_- < \lambda_0.
\end{eqnarray}

Henceforth $\lambda_{\pm}$ are arbitrary (with regard to the above
inequalities) constants.

The quantity $\lambda_0\le 2(u_2^+ + \lambda_+)$ is defined by the
condition that the velocity $2(u_1+u_2^+)$ of the wave $S_2$ (see
(\ref{second})) must be greater than the velocity $2(u_1^- - u_2)$
of the wave $S_1$ (see (\ref{third})). We obtain the inequality
$2u_2^+ + \lambda_+ - \lambda_0 \ge 0$. The maximum $\lambda_0$
satisfying this condition is $2u_2^+ + \lambda_+$. The set of
points $u^-$ satisfying (\ref{u^-(1)}) under the condition
$\lambda_0 = 2u_2^+ + \lambda_+$ coincides with region 1 in Fig.
\ref{R11N}.

{\bf Region  2}. Suppose that the solution of the Riemann problem
is constructed from the fast shock $\hat S_2$, $u\to u^+$, and the
slow shock $S_1$, $u^-\to u$. This is possible when
$$
  u = (u_1^+ + \lambda_+, u_2^+ - \lambda_+), \qquad   \lambda_+ > 2 u_2^+.
$$
In this case,
\begin{equation}
\label{u^-(2)}
  u^- = (u_1^+ + \lambda_+ + \lambda_-, u_2^+ - \lambda_+ + \lambda_-), \qquad
  0 < \lambda_- < \lambda_0.
\end{equation}

According to (\ref{third}), the velocity of the wave $\hat S_2$ is
$2(u_1 - u_2^+) = 2(u_1^+ + \lambda_+ - u_2^+)$. The velocity of
the wave $S_1$ is calculated with the use of (\ref{second}):
$2(u_1^- + u_2) = 2(u_1^+ + \lambda_- + u_2^+)$. We have the
inequality $\lambda_0 = \lambda_+ - 2 u_2^+$.

The set of points $u^-$ defined by inequalities (\ref{u^-(2)})
yields region 2 in Fig. \ref{R11N}.

{\bf Region 3}. Consider the situation where the solution of the
Riemann problem is given by a combination of the fast shock $\hat
S_2$ (transition $u\to u^+$) and the slow rarefaction wave $R_1$
(transition $u^-\to u$). Then
\begin{eqnarray}
\label{u(3)}
  u = (u_1^+ + \lambda_+, u_2^+ - \lambda_+), \qquad \lambda_+ > 2 u_2^+, \\
\label{u^-(3)}
  u^- = (u_1^+ + \lambda_+ - \lambda_-, u_2^+ - \lambda_+ - \lambda_-), \qquad \lambda^- > 0.
\end{eqnarray}
The set of points (\ref{u^-(3)}) defines region 3 in Fig.
\ref{R11N}.

Let us verify the inequality between the velocities of waves. The
velocity of the wave $\hat S_2$ is $2(u_1 - u_2^+) = 2(u_1^+ +
\lambda_+ - u_2^+)$. The velocity of the rarefaction wave at the
right end $u$ can be calculated by formulas
(\ref{Ws})--(\ref{Wf}). To this end, we have to take into account
the inequality $u_2<0$, so that the velocity $2(u_1+u_2)=2(u_1^+ +
u_2^+)$ of the slow wave describes the second equality (\ref{Wf}),
in which one should take $u$ instead of $u^+$. We obtain
inequality (\ref{u(3)}).

{\bf Region 4}. In region 4, the solution represents a complex
wave consisting of the fast rarefaction wave $R_2$ ($u\to u^+$),
the Jouguet shock $S$ ($B_u\to u$), and the slow rarefaction wave
$R_1$ ($u^- \to B_u$). The point $B_u$ defining the Jouguet wave
satisfies the equality
$$
  u - B_u = 2 u_2 (1,-1).
$$
According to (\ref{fast_rare})--(\ref{slow_rare}),
$$
  u^+ - u = \lambda_+ (1, 1), \quad
  B_u - u^- = \lambda_- (1, 1), \qquad
  \lambda_\pm > 0.
$$
Here the difference $B_u - u^-$ is calculated with regard to the
fact that the second coordinate of the point $B_u$ is negative.

The velocity of the Jouguet wave $S$ is calculated in Section
\ref{sec:Jouguet} and is $2u_1 + 2u_2$. It coincides with the
minimum velocity at the rarefaction wave $R_2$ (the characteristic
velocity behind the fast rarefaction wave) and the maximum
velocity at the rarefaction wave $R_1$ (the characteristic
velocity ahead of the slow rarefaction wave). Thus,
\begin{eqnarray}
\nonumber
     u
 &=& (u_1^+ - \lambda_+, u_2^+ - \lambda_+), \quad
     B_u
\, =\, \big(u_1^+ - \lambda_+ + 2(u_2^+ - \lambda_+), - u_2^+ + \lambda_+\big), \\
\label{u-}
     u^-
 &=& \big(u_1^+ + 2u_2^+ - 3\lambda_+ - \lambda_-, - u_2^+ + \lambda_+ - \lambda_- \big).
\end{eqnarray}
The set of points $u^-$ satisfying (\ref{u-}) for $0 < \lambda_+ <
u_2^+$ and $0 < \lambda_-$ is region 4.

{\bf Region 5}. The solution $R_2 RR_1$ of the Riemann problem in
region 5 consists of three rarefaction waves. The first is the
fast wave $R_2$ ($u\to u^+$), where $u$ lies on the axis $u_1$.
This wave is followed by the special rarefaction wave $R$ ($\hat
u\to u$), $\hat u_2=0$. The last wave is another rarefaction wave
$R_1$ ($u^-\to\hat u$), which is denoted here as a slow wave,
although, as discussed at the end of Section \ref{sec:RH}, there
is no point in speaking of fast or slow waves in the situation
where the right asymptotics of the wave lies on the axis $u_1$.

The coordinates of the points $u$, $\hat u$, and $u^-$ are as
follows:
\begin{eqnarray}
\nonumber
&   u
  = (u^+_1 - u^+_2,0), \quad
    \hat u
  = (u^+_1 - u^+_2 - \lambda_+,0), &  \\
\label{u^-(5)}
&   u^-
  = (u^+_1 - u^+_2 - \lambda_+ - \lambda_-, \pm\lambda_-), \qquad
    \lambda_+,\lambda_- > 0.
\end{eqnarray}
The region formed by the points $u^-$ (\ref{u^-(5)}) is region 5
in Fig. \ref{R11N}.

{\bf Region 6}. In region 6, the solution is represented by a
combination of the fast $R_2$ ($u\to u^+$, $u_2
> 0$) and slow $R_1$ ($u^-\to u$) rarefaction waves. The points $u$
and $u^-$ have coordinates
\begin{eqnarray}
\nonumber
&   u
 = (u^+_1 - \lambda_+,u^+_2 - \lambda_+), \qquad
   0 < \lambda_+ < u^-, &  \\
\label{u^-(6)}
&   u^-
  = (u^+_1 - \lambda_+ - \lambda_-, u^- - \lambda_+ + \lambda_-), \qquad
    \lambda_- > 0.
\end{eqnarray}
The set of points $u^-$ (\ref{u^-(6)}) coincides with region 6 in
Fig. \ref{R11N}.

{\bf Region 7}. In region 7, the solutions of the Riemann problem
are given by a combination of the fast shock $S_2$ ($u\to u^+$)
and the slow rarefaction wave $R_1$ ($u^-\to u$). In this
situation,
\begin{eqnarray}
\nonumber
&   u
 = (u^+_1 + \lambda_+,u^+_2 + \lambda_+), \qquad
   \lambda_+ > 0, &  \\
\label{u^-(7)}
&   u^-
  = (u^+_1 + \lambda_+ - \lambda_-, u^- + \lambda_+ - \lambda_-), \qquad
    \lambda_- > 0.
\end{eqnarray}
Region 7 is formed by the points $u^-$ (\ref{u^-(7)}).

{\bf Region 8}.  Finally, region 8 corresponds to solutions in the
form of a combination of the fast rarefaction wave  $R_2$ ($u\to
u^+$), $u_2 > 0$ and the slow shock $S_2$ ($u^-\to u$). In this
case,
\begin{eqnarray}
\nonumber
&   u
 = (u^+_1 - \lambda_+,u^+_2 - \lambda_+), \qquad
   0 < \lambda_+ < u_2^+, &  \\
\label{u^-(8)}
&   u^-
  = (u^+_1 - \lambda_+ + \lambda_-, u^- - \lambda_+ - \lambda_-), \qquad
    0 < \lambda_- < \lambda_+ - u^+_2.
\end{eqnarray}
The points $u^-$ (\ref{u^-(8)}) fill the triangle 8.

\subsection{The case of $\mu_1 > \mu_2$}

In the case of $\mu_1>\mu_2$, undercompressive shocks are
possible. In this case, the points $u^-$ defining the structure of
a slow wave fill only the subinterval $AD$, where the point $D$ is
defined by equality (\ref{D=()}), rather than the whole interval
$AB$. Similarly, not all the points $u^-$ lying on the ray of the
Hugoniot locus going downward right from $B$ define fast shocks
with structure. This property is satisfied only by the points
lying below the point $E$ (\ref{E=()}). These facts lead to the
necessity to include undercompressive shocks in the solutions of
the Riemann problem.

\begin{figure}[h]
\centerline{\psfig{figure=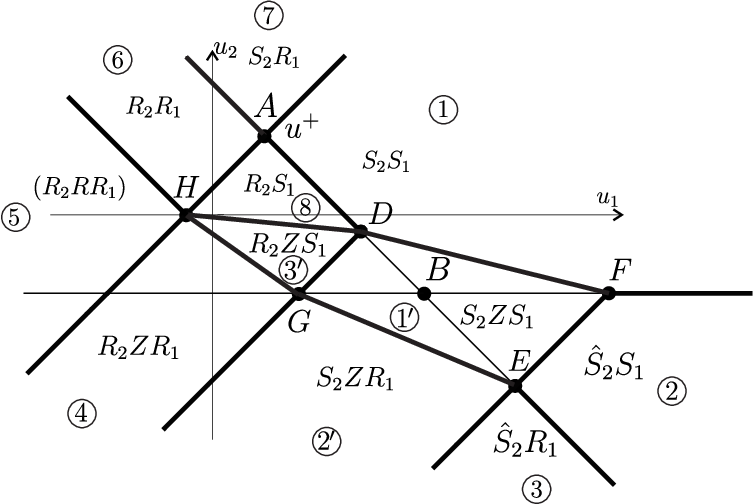,height=7cm}} \caption{Solution
of the Riemann problem in the case of $\mu_1>\mu_2$. }
\label{R22N}
\end{figure}

The solutions in regions 5, 6, and 7 are the same as the solutions
in the same regions in the case of $\mu_1\le \mu_2$.

{\bf Region 1}. We begin with the situation where the solution of
the Riemann problem consists of the fast shock $S_2$ (transition
$u\to u^+$) followed by the slow shock $S_1$ (transition $u^-\to
u$). The coordinates of the points $u$ and $u^-$ are as follows:
\begin{eqnarray}
\nonumber
      u
  &=& (u_1^+ + \lambda_+, u_2^+ + \lambda_+), \qquad \lambda_+ > 0, \\
\label{u^-(1+)}
      u^-
  &=& (u_1^+ + \lambda_+ + \lambda_-, u_2^+ + \lambda_+ - \lambda_-), \qquad
      0 < \lambda_- < \lambda_0.
\end{eqnarray}
The quantity $\lambda_0\le 2(u_2^+ + \lambda_+)$ is determined by
the following two facts.
\smallskip

(a) the velocity $2(u_1+u_2^+)$ of the wave $S_2$ (see
(\ref{second})) must be greater than the velocity $2(u_1^- - u_2)$
of the wave $S_1$ (see (\ref{third})),

(b) the wave $S_1$ must have a structure; i.e., the point $u^-$
must belong to the interval $(A_u,D_u)$, where $A_u=u$ and $D_u$
is defined by equality (\ref{D=()}) with $u^+=u$.
\smallskip

Condition (a) yields the inequality $\lambda_0 \le 2 u_2^+ +
\lambda_+$. According to (\ref{D=()}), condition (b) implies
$\lambda_0 \le (1 + m) u_2$. Thus, region 1 consists of points
$u^-$ (\ref{u^-(1+)}), where
$$
  \lambda_+ > 0, \quad
  0 < \lambda_- < \min \{2 u_2^+ + \lambda_+ , (1+m) (u_2^+ + \lambda_+)\}.
$$

{\bf Region 2}. Suppose that a solution of the Riemann problem is
constructed from the fast shock $\hat S_2$ ($u\to u^+$) and the
slow shock $S_1$ ($u^-\to u$). This is possible when
$$
  u = (u_1^+ + \lambda_+, u_2^+ - \lambda_+), \qquad
  \lambda_+ > (1 + 1/m) u_2^+.
$$
In this case,
\begin{equation}
\label{u^-(2+)}
  u^- = (u_1^+ + \lambda_+ + \lambda_-, u_2^+ - \lambda_+ + \lambda_-), \qquad
  0 < \lambda_- < \lambda_0.
\end{equation}

The quantity $\lambda_0$ is defined by the following two
conditions:
\smallskip

(a) the velocity of the wave $\hat S_2$ is greater than the
velocity of $S_1$,

(b) the wave $S_1$ has a structure.
\smallskip

According to (\ref{third}), the velocity of the wave $\hat S_2$ is
$2(u_1 - u_2^+)=2(u_1^+ + \lambda_+ - u_2^+)$. The velocity
$2(u_1^- + u_2) = 2(u_1^+ + \lambda_- + u_2^+)$ of the wave $S_1$
is calculated with the use of (\ref{second}). We have the
inequality $\lambda_- < \lambda_+ - 2u_2^+$.

A quantitative version of condition (b) follows from Proposition
\ref{rem:D}: $u^-$ belongs to the interval $(u,D_u)$ and $B_u = u
- (1+m) u_2 (1,1)$. In other words,
$$
  0 < \lambda_- < (1 + m)(\lambda_+ - u_2^+).
$$
The set of points $u^-$ (\ref{u^-(2+)}) under the conditions
$$
  \lambda_+ > (1 + 1/m) u_2^+, \quad
  0 < \lambda_- < \min\{\lambda_+ - 2u_2^+, (1+m)(\lambda_+ - u_2^+)\}
$$
defines region 2 in Fig. \ref{R22N}. The point $F$ has coordinates
$(u_1^+ + 2m^{-1} u_2^+, -u_2^+)$.

{\bf Region 3}. The Riemann  problem can also have a solution
$\hat S_2 R_1$. The corresponding fast shock $S_2$ ($u\to u^+$)
and the slow rarefaction wave $R_1$ ($u^-\to u$) are such that
\begin{eqnarray}
\nonumber
    u &=& (u_1^+ + \lambda_+,u_2^+ - \lambda_+), \qquad  \lambda_+ > (1 + 1/m) u_2^+, \\
\label{u^-(5'+)}
  u^- &=& (u_1^+ + \lambda_+ - \lambda_-,u_2^+ - \lambda_+ - \lambda_-), \qquad \lambda_- > 0.
\end{eqnarray}
The velocity $W_2$ of the wave $\hat S_2$ is $W_2 = 2(u_1 - u_2^+)
= 2(u_1^+ + \lambda_+ - u_2^+)$. The maximum velocity of points on
the wave $R_1$ is as follows: $C_1 = 2(u_1 + u_2) = 2(u_1^+ +
u_2^+)$. The inequality $C_1 < W_2$ has the form $\lambda_+
> 2u_2^+$. The set of points $u^-$ (\ref{u^-(5'+)}) coincides with region $4'$ in Fig. \ref{R22N}.

{\bf  Region} $\bf 1'$. In region $1'$, the solution consists of a
sequence of three waves: the fast shock $S_2$ (transition $u\to
u^+$), the undercompressive shock $Z$ (transition $u^\times\to
u$), and the slow shock $S_1$ (transition $u^-\to u^\times$). In
this case,
\begin{eqnarray}
\label{u1'1}
\!\!\!\!\!\!
      u
  &=& (u_1^+ + \lambda_+, u_2^+ + \lambda_+), \qquad  \lambda_+ > 0, \\
\label{u1'2}
\!\!\!\!\!\!
      u^\times
  &=& (W_\times - u_1^+ - \lambda_+, -u_2^+ - \lambda_+ ), \qquad
      W_\times
   =  2(u_1^+ + \lambda_+ + m(u_2^+ + \lambda_+)).
\end{eqnarray}
The position of the point $u^\times$ and the velocity $W_\times$
are calculated with regard to (\ref{4cp}) and Proposition
\ref{prop:sc}. The point $u^-$ is determined from the equation
\begin{equation}
\label{u^-(1'+)}
  u^- = (u_1^+ + \lambda_+ + 2m(u_2^+ + \lambda_+) + \lambda_-, -u_2^+ - \lambda_+ + \lambda_-).
\end{equation}
The velocities of the waves $S_2$ and $S_1$ satisfy the equalities
\begin{eqnarray*}
      W_2
  &=& 2(u_1 + u_2^+)
   =  2(u_1^+ + \lambda_+ + u_2^+), \\
      W_1
  &=& 2(u_1^- + u_2^\times)
   =  2(u_1^+ + \lambda_+ + 2m(u_2 + \lambda_+) + \lambda_- - u_2^+ - \lambda_+).
\end{eqnarray*}
The inequality $W_1 < W_\times < W_2$ takes the form
\begin{equation}
\label{(1')}
  m\lambda_+ < (1 - m) u_2^+, \quad
   \lambda_- < (1 - m) (u_2^+ + \lambda_+).
\end{equation}

The set of points (\ref{u^-(1'+)}) with conditions (\ref{(1')}) is
the rectangle $GDFE$, where $D = (u_1^+ + 2m u_2^+, - u^+_2)$.

{\bf Region} ${\bf 2'}$. Consider the region in which the solution
of the Riemann problem has the form $S_2 Z R_1$, where $S_2$ is
the fast shock $u\to u^+$, $Z$ is the undercompressive shock
$u^\times\to u$, and $R_1$ is the slow rarefaction wave $u^-\to
u^\times$. The points $u$ and $u^\times$ are calculated by
formulas (\ref{u1'1})--(\ref{u1'2}). For the point $u^-$ we have
\begin{equation}
\label{u^-(2'+)}
  u^- = \big(u_1^+ + \lambda_+ + 2m(u_2^+ + \lambda_+) - \lambda_-, -u_2^+ - \lambda_+ - \lambda_-\big), \quad
  \lambda_- > 0.
\end{equation}
The maximum velocity of particles in the wave $R_1$ is
$$
  C_1 = 2(u_1^- + u_2^-)
      = 2(u_1^+ - u_2^+ + 2m(u_2^+ + \lambda_+) - 2\lambda_-).
$$
The inequality $C_1 < W_\times < W_2$ ($W_\times$ and $W_2$ are
defined in region ${\bf 1'}$) takes the form
\begin{equation}
\label{(2')}
  m\lambda_+ < (1 - m) u_2^+, \quad
   \lambda_- > (m - 1) (u_2^+ + \lambda_+).
\end{equation}
Since $m-1<0$, $u_2^+ > 0$, and $\lambda_\pm > 0$, the second
inequality in (\ref{(2')}) holds automatically. The region formed
by the points $u^-$ (\ref{u^-(2'+)}) under conditions (\ref{(2')})
is denoted by number $ 2'$ in Fig. \ref{R22N}.

{\bf Region} ${\bf 3'}$. Suppose that the solution of the Riemann
problem has the form $R_2 Z S_1$, where $R_2$ is the fast
rarefaction wave $u\to u^+$, $Z$ is then undercompressive shock
$u^\times\to u$, and $S_1$ is the slow shock $u^-\to u^\times$.
Then
\begin{eqnarray}
\label{u3'}
\!\!\!\!\!\!
      u
  &=& (u_1^+ - \lambda_+, u_2^+ - \lambda_+), \qquad  0 < \lambda_+ < u_2^+, \\
\label{ux3'}
\!\!\!\!\!\!
      u^\times
  &=& (W_\times - u_1^+ + \lambda_+, -u_2^+ + \lambda_+), \qquad
      W_\times
   =  2(u_1^+ - \lambda_+ + m(u_2^+ - \lambda_+)), \\
\label{u^-(3'+)}
\!\!\!\!\!\!
      u^-
  &=& \big(u_1^+ - \lambda_+ + 2m(u_2^+ - \lambda_+) + \lambda_-, -u_2^+ + \lambda_+ + \lambda_-\big), \quad
  \lambda_- > 0.
\end{eqnarray}
The minimum velocity $C_2$ of particles in the wave $R_2$ and the
velocity $W_1$ of the wave $S_1$ are
\begin{eqnarray}
\label{W23'}
  C_2 &=& 2(u_1 + u_2) = 2(u_1^+ + u_2^+ - 2 \lambda_+), \\
\nonumber
  W_1 &=& 2(u_1^- + u_2^\times) = 2\big(u_1^+ + (2m-1)u_2^+ - 2m\lambda_+ + \lambda_- \big).
\end{eqnarray}
The inequality $W_1 < W_\times < C_2$ takes the form
\begin{equation}
\label{(3')}
   \lambda_+ < u_2^+, \quad
   \lambda_- < (1 - m) (u_2^+ - \lambda_+).
\end{equation}
The region formed by the points $u^-$ (\ref{u^-(3'+)}) under
conditions (\ref{(3')}) is the triangle $HDG$,
$$
  H = (u_1^+ - u_2^+, 0).
$$

{\bf Region 4}. Consider the situation where the solution of the
Riemann problem is represented as $R_2 Z R_1$. The fast
rarefaction wave $R_2$ ($u\to u^+$), the undercompressive shock
$Z$ ($u^\times\to u$), and the slow rarefaction wave $R_1$
($u^-\to u^\times$) are defined by conditions
(\ref{u3'})--(\ref{ux3'}):
\begin{equation}
\label{u^-(4'+)}
      u^-
   =   \big(u_1^+ - \lambda_+ + 2m(u_2^+ - \lambda_+) - \lambda_-, -u_2^+ + \lambda_+ - \lambda_-\big), \quad
  \lambda_- > 0.
\end{equation}
The minimum velocity $C_2$ of particles in the wave $R_2$
satisfies equality (\ref{W23'}). The velocity of the
undercompressive shock is shown in (\ref{ux3'}). The maximum
velocity of particles in the wave $R_1$ is determined from
(\ref{Wf}) with regard to the inequality $u_2^\times < 0$:
$$
    C_1
  = 2(u_1^\times + u_2^\times)
  = 2\big(u_1^+ + (2m-1)u_2^+ - 2m\lambda_+ \big).
$$
The inequality $C_1 < W_\times < C_2$ reduces to the form
$\lambda_+ < u_2^+$. The set of points (\ref{u^-(4'+)}) is region
$4'$ in Fig. \ref{R22N}.

{\bf Region 8}. It remains to consider the region in which the
Riemann problem has solutions in the form of a combination of the
fast rarefaction wave $R_2$ ($u\to u^+$, $u_2 > 0$) and the slow
shock $S_1$ ($u^-\to u$). In this case,
\begin{eqnarray}
\nonumber
&   u
 = (u^+_1 - \lambda_+,u^+_2 - \lambda_+), \qquad
   0 < \lambda_+ < u_2^+, &  \\
\label{u^-(6'+)}
&   u^-
  = (u^+_1 - \lambda_+ + \lambda_-, u^- - \lambda_+ - \lambda_-), \qquad
    0 < \lambda_- < \lambda_0.
\end{eqnarray}
To calculate $\lambda_0$, we have to take into account the
following two conditions:
\smallskip

(a) The minimum velocity $C_2=2(u_1+u_2)=2(u_1^+ + u_2^+ -
2\lambda_+)$ on the wave $R_2$ must be greater than the velocity
$W_1=2(u_1^- - u_2)=2(u_1^+ + \lambda_- - u_2^+)$ of the wave
$S_1$.

(b) The wave $S_1$ must have a structure: $\lambda_- < (1+m)(u_2^+
- \lambda_+)$.
\smallskip

Hence we obtain $\lambda_0 = (1+m)(u_2^+ - \lambda_+)$. The points
$u^-$ (\ref{u^-(6'+)}) fill the triangle $ADH$.

\section{Conclusions}
For a regularized $2\times2$ system ({\ref{reg}}, {\ref{Q=}}) of
hyperbolic conservation laws, we have distinguished the parts of
the Hugoniot locus that correspond to shocks with structure. We
have shown that not all shocks have a structure. We have
highlighted parts of the Hugoniot locus that correspond to shocks,
Jouguet waves, undercompressive shocks, and overcompressive
shocks. We have shown that undercompressive shocks are possible
only in the case of $\mu_1>\mu_2$.

We have constructed a solution to the Riemann problem. In the case
of $\mu_1\leqslant\mu_2$, all shocks have a structure and that
there are no undercompressive shocks. The solution represents a
sequence of shocks and rarefaction waves. In the case of $\mu_1>
\mu$, not all shocks have a structure, and there exist
undercompressive shocks. In this case, the sequence of waves
representing the solution of the Riemann problem contains an
undercompressive shock in some regions.

We have proved the uniqueness of the solution of the Riemann
problem.

\section*{Acknowledgments}

This work was supported by the Russian Science Foundation under grant no. 25-11-00114.

\end{document}